\documentclass[reprint,amsmath,amssymb,aps,superscriptaddress,onecolumn]{revtex4-1}

\usepackage{amssymb}
\usepackage{amsmath}
\usepackage{graphicx,color}
\usepackage{dcolumn}
\usepackage{bm}
\usepackage{epsfig,color}
\usepackage{epstopdf}
\usepackage[colorlinks=true,linkcolor=black,citecolor=blue]{hyperref}
\usepackage{float}

\begin{document}

\date{\today}

\title{Fractional Dynamics and Modulational Instability in Long-Range Heisenberg Chains}

\author{Mbetkwe Youwa Laetitia}
\affiliation{Pure Physics Laboratory: Group of Nonlinear
Physics and Complex Systems, Department of Physics, Faculty of
Science, University of Douala, Box 24157, Douala, Cameroon}

\author{Jean Pierre Nguenang}
\affiliation{Pure Physics Laboratory: Group of Nonlinear
Physics and Complex Systems, Department of Physics, Faculty of
Science, University of Douala, Box 24157, Douala, Cameroon}

\author{Paul Andr\'e Paglan}
\affiliation{Pure Physics Laboratory: Group of Nonlinear
Physics and Complex Systems, Department of Physics, Faculty of
Science, University of Douala, Box 24157, Douala, Cameroon}
\affiliation{National Committee for Development of Technologies (NCDT), Ministry of Scientific Research
              and Innovation - P.O. Box 1457 Yaounde, Cameroon}

\author{Thierry Dauxois}
\affiliation{ENS de Lyon, CNRS,
Laboratoire de Physique, F-69342 Lyon, France}

\author{Andrea Trombettoni}
\affiliation{Department of Physics, University of Trieste, Strada Costiera
  11, I-34151 Trieste, Italy}
\affiliation{SISSA, Via Bonomea 265, I-34136
  Trieste, Italy}
\affiliation{INFN, Sezione di Trieste, I-34151
Trieste, Italy}

\author{Stefano Ruffo}
\affiliation{SISSA, Via Bonomea 265, I-34136 Trieste, Italy}
\affiliation{INFN, Sezione di Trieste, I-34151 Trieste, Italy}
\affiliation{Istituto dei Sistemi Complessi, Consiglio Nazionale
delle Ricerche, via Madonna del Piano 10, I-50019 Sesto Fiorentino,
Italy}

\begin{abstract}
We study the \textcolor{black}{effective} dynamics of ferromagnetic spin chains in presence of long-range interactions.
We consider the Heisenberg Hamiltonian in one dimension for which the spins are coupled through
power-law long-range exchange interactions with exponent $\alpha$. \textcolor{black}{We add to the Hamiltonian} an anisotropy in the $z$-direction. 
In the framework of a semiclassical approach, we use the Holstein-Primakoff transformation
to derive an effective long-range discrete nonlinear Schr\"odinger equation. 
We then perform the continuum limit and we obtain a {\em fractional} nonlinear Schr\"odinger-like equation.
Finally, we study the modulational instability of plane-waves in the continuum limit and we prove that,
at variance with the short-range case, plane waves are modulationally unstable for 
$\alpha < 3$.
\textcolor{black}{We also study the dependence of the modulation instability growth rate and critical wave-number
on the parameters of the Hamiltonian and on the exponent $\alpha$.}


\medskip
\textbf{\emph{Keywords}}: \emph{Heisenberg spin chains, long-range interactions, fractional 
equations, modulational instability.}
\end{abstract}

\maketitle
\section{Introduction}\label{sec:I}

\textcolor{black}{Simplified models of magnetic systems, like the Ising model, have allowed the understanding of 
complex magnetic phenomena and the theoretical interpretation of phase transitions~\cite{parisi1986}. Specifically, the study of low dimensional spin systems has been at the center of a constant attention in the field of magnetic materials~\cite{mattis1986,majlis2007,lacroix2011}.
Applications of models of magnetism cover a very wide range
of phenomena, from classical and quantum transport to nuclear magnetic resonance~\cite{a1}. An important role in the study of magnetic systems is played by the description of strong correlations and of the nonlinearities induced by inter-spin interactions \cite{a2,a3}. 
On the other hand, the effects of long-range interactions, where each unit is coupled to all others, motivated a remarkable activity in the last 
decades~\cite{a16,campa2014,defenu2020}. 
First studies of models of classical spins with long-range interactions
date back to Refs.~\cite{dyson1,dyson2}. A typical interaction which is relevant for a number of systems, ranging from gravitational ones to dipolar magnets and gases, is provided by the power-law decay ${1}/{r^{\alpha}}$, where $r$ is the distance among constituents. It is well known that a first criterion to determine the long-range nature of a system with power-law decaying interactions is based on the comparison of $\alpha$ with 
the space dimension $d$. If $ \alpha < d$ and if the system is homogeneous, then it displays a diverging energy density.}
Thus, in order to obtain a well defined thermodynamic limit, the so-called Kac rescaling of the energy is required. In the following, 
we will refer to this region as the non-additive or strong long-range region. If $\alpha>d$, the energy of the system is additive 
and it is useful to introduce a value of $\alpha$, which we denote by $ \alpha^*$, such that for $ \alpha>\alpha^*$ the system 
behaves at criticality as a short-range system~\cite{sak1973}. Since $ \alpha^*>d$, there is a region of values of $\alpha$, given 
by $d<\alpha<\alpha^*$, in which the behavior of the system significantly differs from the one of the same system with short-range 
interactions, although the energy is additive: such a region is the weak long-range region. The actual value  of $\alpha^*$ depends
on the specific model and the dimension~$d$. As an example, for classical $O(n)$ models in $d = 1$, it is known that $\alpha^* = 2 $, see 
e.g.~\cite{thouless1969,a17,a18}. These studies have been also performed for quantum spin systems with long-range 
couplings~\cite{dutta2001,defenu2017}. 

A common \textcolor{black}{feature} for systems with long-range interactions is the behaviour of the dispersion relation as $\propto k^\sigma$  
where $\sigma=\alpha-d$. \textcolor{black}{This behaviour implies, in many cases, that the low-energy effective dynamics can be described by a fractional 
differential equation~\cite{a15}, which involves derivatives and/or integrals of non-integer order. Such equations are used to describe anomalous
kinetics, anomalous transport phenomena, etc.~\cite{a19}.} Fractional dynamics is a field of study which investigates the behavior of microscopic
and macroscopic systems that are either characterized by power-law non locality, or by power-law 
long term memory. \textcolor{black}{Recently, models of coupled nonlinear oscillators with long-range couplings have been introduced with the aim of understanding
the role played by nonlocal interactions in several phenomena: relaxation to equilibrium~\cite{Kevrekidi20}, anomalous transport~\cite{Benenti21},
localized solutions~\cite{Korabel20}.} Among various possible examples, we can mention the Fermi-Pasta-Ulam-Tsingou model~\cite{Bountis20} with long-range couplings~\cite{a6,a7}, for which one can derive explicitly an effective fractional equation, the fractional Boussinesq equation, whose 
solutions depend on the long-range exponent $\alpha$. 

\textcolor{black}{A natural extension of these studies is to consider} other models, in particular those relevant for the study of magnetism. In this paper, we investigate the Heisenberg model with long-range couplings \cite{Frohlich,Joyce} and we derive \textcolor{black}{for the first time} in the large-$S$ limit, $S$ being the value of the spin, a fractional nonlinear 
Schr\"odinger (FNLS) equation for the effective dynamics. To this aim, we proceed by mapping the Heisenberg spin chain with long-range interactions
into a continuum equation with the Riesz fractional derivatives using the Holstein-Primakoff transformation. The final equations of motion describe the dynamics of the Heisenberg model with long-range couplings in the continuum limit. In order to show an application of the equations that we obtain, we study modulational instabilities and we discuss how the modulationally unstable and  stable regions depend on the range of the power-law 
exponent $\alpha$. Modulational instabilities were introduced and studied in lattice models~\cite{kivshar1992,daumont1997,trombettoni2001} to study the localization of energy and the stability of dynamical regimes~\cite{dauxois2006}. 
In particular, the modulational instabilities of the discrete nonlinear Schr\"odinger equations with long-range couplings were
considered in Ref.~\cite{a20}. Here, the derivation of the effective fractional equation allows for a qualitative determination 
of modulationally stable regions in the space of parameters.  

The paper is organized as follows. In Section II, we derive the effective
Hamiltonian using bosonic operators. In Section III,
we derive the FNLS equation and the corresponding integro-differential equation valid in the continuum limit. 
In Section IV, we study the modulational instability of plane-waves excitations. In Section V, we discuss the results and provide an 
outlook on future directions of investigation. In the Appendices, we present some details of the analytical computations leading to the results discussed in the main 
text.

\section{The model and the Holstein-Primakoff transformation}
In order to study the dynamics of the one-dimensional (1D) Heisenberg spin chain for
spins $\vec S=(S_x,S_y,S_z)$ located on the sites of a chain, let us start by considering the Hamiltonian
\begin{equation}
H =  -A \sum_{n = -N}^{N} {\left( {S_n^z } \right)^2 } - \sum_{-N \leq n<m \leq N} {J_{m,n} \vec S_n  \cdot \vec S_m }  
 \label{eq1}
\end{equation}
where the indices $n$ and $m$ denote the sites of the lattice and $N$ is the total
number of spins. $S$ is the value of the spin. In the first term of the Hamiltonian, $A$ represents  the anisotropy parameter and
in the second term,  the exchange interaction $J_{m,n}$ is supposed to decay
algebraically with distance as a power-law  
\begin{equation}\label{eq2}
J_{m,n}  = \frac{J}{\left| {m - n} \right|^\alpha} . 
\end{equation}
We consider a ferromagnetic chain ($J>0)$ \textcolor{black}{and choose $\alpha$ in the range $1<\alpha<3$, because for smaller values, $0<\alpha \leq 1$, the Hamiltonian diverges}. This latter is the non additive strong long-range region, we
therefore limit in this paper to analyze the weak long-range region. For larger values,
$\alpha >3$, the system in the low-energy limit is expected to display a short-range behavior, which is 
exactly reached only when $\alpha \to \infty$.

\textcolor{black}{The cross product of the spins at two different sites can be rewritten in terms of} annihilation and creation operators as follows
\begin{equation}\label{eq3}
\vec S_n \, \cdot \vec S_m  = \frac{1}{2}\left( {S_n^ +  S_m^ - + S_n^ -  S_m^ +  } \right) + S_n^z S_m^z.
\end{equation}
We then use  the Holstein-Primakoff transformation for spins operators in terms of bosonic operators in the framework of the
low temperature approximation~\cite{a21,a22,a23} as
\begin{eqnarray}\label{eq4}
S_n^ +   &=& \sqrt 2 \varepsilon \left( {1 - \frac{{\varepsilon ^2}} {4}a_n^ +  a_n  + o\left( {\varepsilon ^4 } \right)} \right)a_n\\
S_n^ -  & =& \sqrt 2 \varepsilon \,a_n^ +  \left( {1 -
\frac{{\varepsilon ^2 }} {4}a_n^ +  a_n  + o\left( {\varepsilon ^4
} \right)} \right)\\
S_n^z   &=& 1 - \varepsilon ^2 a_n^ +  a_n,\label{eq4bis}
\end{eqnarray}
where $\varepsilon=1/\sqrt{S}$.
Therefore, \textcolor{black}{in the limit of large $S$,} substituting Eqs.~(\ref{eq4}-\ref{eq4bis}) into Eq.~(\ref{eq3}), we get
\begin{eqnarray}
\vec S_n  \cdot \,\vec S_m  &=& 1 + \varepsilon ^2 \left(a_n a_m^ +   + a_n^ +  a_m  - a_n^ +  a_n  - a_m^+ a_m \right) \nonumber\\
 &&- \frac{\varepsilon ^4}{4}\left( a_n^ +  a_n a_n a_m^ +   + a_n a_m^ +  a_m^ +  a_m  + a_n^ +
 a_n^ +  a_n a_m  + a_n^ +  a_m^ +  a_m a_m  - 4a_n^ +  a_n a_m^ +  a_m  \right) +o\left(\varepsilon ^6\right), \label{eq5}
 \end{eqnarray}
 while
 \begin{equation}\label{eq6}
\left( {S_n^z } \right)^2  = 1 - 2\varepsilon ^2 a_n^ +  a_n  +
\varepsilon ^4 a_n^ +  a_n a_n^ +  a_n+o\left(\varepsilon ^6\right).
 \end{equation}
Hamiltonian  (\ref{eq1}) can thus be rewritten in terms of bosonic creation and annihilation operators as
 \begin{eqnarray}\label{eq7}
H_1 &=&  - \sum_{-N \leq n<m\leq N} J_{m,n} \biggl\{ F \left( a_n a_m^ +   + a_n^+a_m  - a_n^+a_n - a_m^+a_m  \right) 
- G\biggl(  a_n^+a_n a_n a_m^+   + a_n a_m^+a_m^ +  a_m  + a_n^+a_n^+a_n a_m  \nonumber \\
 && \hskip 4truecm + a_n^+a_m^+a_m a_m  - 4a_n^+a_n a_m^+a_m    \biggr) \biggr\}
  -\sum\limits_{n =  {\color{black}{-N} }}^{ {\color{black}{N}}} \left(  - T a_n^ +  a_n  +I a_n^ +  a_n a_n^ +  a_n \right)+o\left(\varepsilon ^6\right) ,
\end{eqnarray}
\textcolor{black}{where} $H_1=H+NA+ \sum_{-N \leq n<m\leq N} J_{m,n}$ and the coupling constants are $F=\varepsilon ^2$, $G={{F ^2 }}/{4}$, $T=2FA$, and $I=AF ^2$. Equation~(\ref{eq7}) is a bosonic, Bose-Hubbard-like, Hamiltonian which \textcolor{black}{will be used in the next Section}.

\section{Equations of motion}
The equations of motion are obtained from  the Heisenberg evolution equation for a given bosonic operator $a_n$ as follows
\begin{equation}\label{eq8}
i\hbar \dot a_n  = \left[ {a_n ,H} \right] \,\,, \,\, -N \leq n \leq N
\end{equation}
One has to compute the different commutators arising from the previous equations to get the following nonlinear discrete equation
 for the bosonic operators
\begin{eqnarray}\label{eq9}
i\hbar \dot a_n  &=&  - \sum\limits_{m \ne n} {J_{mn} \left[ { - F (a_n  - a_m ) - G\left( {a_n a_n a_m^ +   + a_m^ +  a_m a_m    + 4a_n^ +  a_n a_m  - 4a_n a_m^ +  a_m } \right)} \right]} \nonumber \\
&&   \hskip 7truecm 
+ 
 Ta_n  - I \left( {a_n a_n^ +  a_n  + a_n^ +  a_n a_n } \right).
\end{eqnarray}

\textcolor{black}{One can write mean-field, classical equations for the dynamics of the expectation values of the bosonic operators. One way to do it is to
use the Glauber coherent states representation which are defined as 
 $a_n^ +  \left| f \right\rangle  = f_n^* \left| f \right\rangle$ and
 $a_n \left| f \right\rangle  = f_n \left| f \right\rangle$
 and apply it for the full equation of motion through}
$\left\langle f \right|i\hbar \mathop {a_n }\limits^. \left| f
\right\rangle  = \left\langle f \right|\left[ {a_n ,H}
  \right]\left| f \right\rangle$.
We then get the following nonlinear equations of motion for the amplitude $f_n$ of the spin chain excitation's 
\begin{equation}\label{eq10}
i\hbar \mathop {f_n }\limits^.  =   \sum\limits_{m \ne n} £
 J_{mn} \left[ {  F (f_n  - f_m ) + G\left( {f_m^* f_n^2  +\left| {f_m } \right|^2 f_m   + 4\left| {f_n } \right|^2 f_m  - 4\left| {f_m } \right|^2 f_n} \right)} \right] + 
 T f_n  - 2I \left| {f_n } \right|^2 f_n . 
\end{equation}
Equation~(\ref{eq10}) is  a discrete cubic nonlinear Schr\"odinger-like  equation which is the subject of our analysis in the following.

\subsection{The continuum limit}

Following the different steps proposed by Tarasov~\cite{a13,Korabel20}, let us define the operation which transforms the above equation for $f_n(t)$ into a continuum medium equation for $f(x, t)$. We assume that $f_n(t)$ are Fourier coefficients of some function $\hat f(k, t)$. Then, we define the field $\hat f(k, t)$ 
on the interval $[-K/2,K/2]$ as
\begin{equation}
\hat f(k, t)= \sum_{n=-\infty}^{+\infty}f_n(t) e^{-ikx_n} = {\cal F}_\Delta\{f_n(t)\}, \label{eqation4}
\end{equation}
where $x_n = n\Delta x$ and $\Delta x = 2\pi/K$ is a distance between oscillators, and conversely
\begin{equation}
f_n(t) = \frac{1}{K}\int _{-K/2}^{+K/2}dk\ \hat f(k, t) e^{ikx_n} = {\cal F}_\Delta^{-1}\{\hat f(k,t)\}.  \label{eqation5}
\end{equation}
These equations define a Fourier transform, which is obtained in the limit $\Delta x \rightarrow 0$ $(K\rightarrow\infty$). In order to perform this limit, let us 
replace the discrete set of functions $f_n(t) = (2\pi/K)f(x_n,t)$ with continuous function of two variables $f(x,t)$, while letting $x_n = n\Delta x = 2\pi n/K \rightarrow x$. Then, change the sum into an integral, and Eqs.~(\ref{eqation4}) and (\ref{eqation5}) become
\begin{eqnarray}
\tilde f(k,t)&=&\int_{-\infty}^{+\infty} dx \ e^{-ikx} f(x,t)= {\cal F}\{f(x,t)\} \\
 f(x,t)&=&\frac{1}{2\pi} \int_{-\infty}^{+\infty} dk\  e^{ikx} \tilde f(k,t)= {\cal F}^{-1}\{\tilde f(k,t)\} .
\end{eqnarray}
Note that $\tilde f(k,t)$ is the Fourier transform of the field $f(x,t)$, and that $\hat f(k,t)$ is the Fourier series  of $f_n(t)$, defined by $f_n(t) = (2\pi /K)f(n\Delta x, t)$.

The map of the discrete model into the continuum one can be defined by the chain of transformations $\hat T={\cal F}^{-1} {\cal L}\ {\cal F}_\Delta$, where ${\cal F}_\Delta$ is 
the Fourier series transform ${\cal F}_\Delta \{f_n(t)\}=\hat f(k,t)$,
the passage to the limit $\Delta x\rightarrow 0$ is represented by the operator ${\cal L}\{\hat f(k,t)\}=\tilde f(k,t)$ and the inverse Fourier transform is ${\cal F}^{-1} \{\tilde f(k,t)\}= f(x,t)$. One has therefore 
\begin{equation}
  f_n(t) 
  \xrightarrow{ {\cal F}_\Delta}      
  \hat f(k,t) 
  \xrightarrow{ {\cal L}} \tilde f(k,t)
  \xrightarrow {{\cal F}^{-1}}    
f(x,t) = \hat T \{  f_n(t) \}.
\end{equation}

After introducing the general formalism needed to perform the continuum limit, let us analyze, using these methods, the discrete description displayed in Eq.~(\ref{eq10}) and derive the corresponding continuum one.  Let us begin by analytically evaluating  each of the terms in this Equation.
Since most of the analytic computations will be performed in the framework of the continuum limit, 
we will consider the limit $N\rightarrow\infty $.

To this end, let consider the first term
\begin{equation}
A_1 = {\sum \limits_{{ \color{black}{-N}}\leq n<m\leq { \color{black}{N}} }
{J_{m,n}.\left( {f_n \textcolor{black}{(t)} - f_m\textcolor{black}{(t)} } \right)}}\label{definitionA1}.
\end{equation}
The computations are performed in Appendix~\ref{AppendixB} in the similar spirit of those in which fractional 
equations are derived in Refs.~\cite{a12,a13,Korabel20,a15,a24,a25,a26}. We obtain
 \begin{equation}\label{eq11}
 A_1 = a_\alpha  \frac{{\partial ^{\alpha-1} 
f(x,t)}}{{\partial \left| x \right|^{\alpha-1}  }}.
\end{equation}
where
 \begin{equation}\label{aalpha}
a_\alpha   = 2J\Gamma (1 - \alpha )\sin \left(\pi \alpha/2 \right)
\end{equation}
where $\Gamma$ is the Euler gamma function and the partial Riesz derivative is defined as
\begin{equation}
-\frac{{\partial ^\alpha u\left( {x,t} \right) }}{{\partial \left| x \right|^\alpha
}} = \frac{1}{{2\pi }}\int\limits_{ - \infty
}^{ + \infty } {dp\left| p \right|^\alpha  \hat u\left( {p,t}
\right)} e^{ipx}. \label{continuumalpha9}
\end{equation}

The following term reads
\begin{eqnarray}\label{eq12}
B  & =&  \sum\limits_{{ \color{black}{-N}}\leq n<m\leq { \color{black}{N}}  } {J_{m,n}
f_m^* f_n^2 }\\
  & = &\left( {\frac{1}{{2\pi }}} \right)^3 \int\limits_{ - \infty }^{ + \infty } {\sum\limits_{{ \color{black}{-N}}\leq n<m\leq { \color{black}{N}}  } {\frac{J}{{\left| {m - n} \right|^{\alpha } }}\tilde{f}^ *  (k,t)e^{ - ikm} } dk\int\limits_{ - \infty }^{ + \infty } {\tilde{f}(k^{'} ,t)e^{ik^{'}n} dk^{'} \int\limits_{ - \infty }^{ + \infty } {\tilde{f}(k^{{''}} ,t)e^{ik^{{''}}n} dk^{{''}} } } } .
\end{eqnarray}
 In the continuum limit  {$ N\rightarrow {\infty}  $}
and setting
$m - n = m^{'}$,  
we get
\begin{eqnarray}\label{eq13}
B &=& \left( {\frac{1}{{2\pi }}} \right)^3 \int\limits_{ - \infty
}^{ + \infty } {\sum\limits_{m^{'} =  - \infty ,m^{'} \ne 0}^{ +
\infty } {\frac{{Je^{ - ikm^{'}} }}{{\left| {m^{'} }
\right|^{\alpha } }}\tilde{f}^ *  (k,t)e^{ - ikn} }
dk\int\limits_{ - \infty }^{ + \infty } {\tilde{f}(k^{'} ,t)e^{ik^{'}
n} dk^{'} \int\limits_{ - \infty }^{ + \infty } {\tilde{f}(k^{{''}} ,t)e^{ik^{{''}} n} dk^{{''}} } } }\\
 &=& \left( {\frac{1}{{2\pi }}} \right)^3 \int\limits_{ - \infty
}^{ + \infty } {\hat J_{\alpha  } ( - k)\tilde{f}^ *
(k,t)e^{ - ikn} dk\int\limits_{ - \infty }^{ + \infty }
{\tilde{f}(k^{'} ,t)e^{ik^{'} n} dk^{'} \int\limits_{ -
\infty }^{ + \infty } {\tilde{f}(k^{''} ,t)e^{ik^{''} n}
dk^{''} } } }.
\end{eqnarray}

It is useful to  introduce the following function 
\begin{equation}
 \hat J_{\alpha  } ( x) = \sum\limits_{m^{'}  =  - \infty ,m^{'}  \ne 0}^{ + \infty }J{\frac{{\displaystyle e^{im^{'} x} }}{{\displaystyle\left| {m^{'}}\right|^\alpha}}}  .
\end{equation}
Next, we use 
the so-called infrared limit approximation, which is helpful to derive the main relation that allows us to pass
from the  discrete medium to the continuum~\cite{Korabel20,a12,a13}. For $ 1< \alpha <3 $, $ \alpha\neq2 $ and $k\longrightarrow0$, 
the fractional power of   $\left| {k}\right| $ is a leading  asymptotic term and 
\begin{equation}\label{eq14}
\hat J_{\alpha  } ( - k) = \hat J_{\alpha  }
(k) \simeq a_\alpha  \left| k \right|^{\alpha-1}   + \hat
J_{\alpha  } (0).
\end{equation}
\textcolor{black}{This allows us} to transform the nonlinear discrete equation  into a fractional differential equation. In the range
$1<\alpha<3  $, provided $ \alpha\neq2 $, we get
\begin{equation}\label{eq15}
B = \left( {\frac{1}{{2\pi }}} \right)^3 a_\alpha  \int\limits_{ -
\infty }^{ + \infty } {\left| k \right|^{\alpha-1}   \tilde{f}^ * (k,t)e^{
- ikn} dk\int\limits_{ - \infty }^{ + \infty } {\tilde{f}(k^{'} ,t)e^{ik^{'} n} dk^{'} \int\limits_{ - \infty }^{
+ \infty } {\tilde{f}(k^{''} ,t)e^{ik^{''} n} dk^{''} } } }+
\hat J_{\alpha  } (0)\left| {f(x,t)} \right|^2f(x,t).
\end{equation}
Therefore, the fractional derivative allows us to write
\begin{equation}\label{eq15_bis}
B = \left( {\frac{1}{{2\pi }}} \right)^3 a_\alpha  \int\limits_{ -
\infty }^{ + \infty } {\left| k \right|^{\alpha-1}   \tilde{f}^ * (k,t)e^{
- ikn} dk\int\limits_{ - \infty }^{ + \infty } {\tilde{f}(k^{'} ,t)e^{ik^{'} n} dk^{'} \int\limits_{ - \infty }^{
+ \infty } {\tilde{f}(k^{''} ,t)e^{ik^{''} n} dk^{''} } } }+
\hat J_{\alpha  } (0)\left| {f(x,t)} \right|^2f(x,t).
\end{equation}
Thereafter, we get
\begin{equation}\label{eq16}
B =  - a_\alpha f^2 (x,t) \frac{{\partial ^{\alpha-1}   f^* (x,t)}}{{\partial
\left| x \right|^{\alpha-1}   }} + \hat J_{\alpha  }
(0)\left| {f(x,t)} \right|^2f(x,t).
\end{equation}
Following similar analytical steps, after 
the computations given in Appendix~\ref{AppendixB},
 the three following terms \textcolor{black}{are derived}
\begin{equation}\label{eq17}
C =  {\sum\limits_{-N\leq n<m\leq N }} 
{J_{mn} f_m \left| {f_m } \right|^2 }= - a_\alpha  \left| {f(x,t)} \right|^2 \frac{{\partial ^{\alpha-1}   f(x,t)}}{{\partial
\left| x \right|^{\alpha-1}   }} + \hat
J_{\alpha  } (0)\left| {f(x,t)} \right|^2f(x,t),
\end{equation}

\begin{equation}\label{eq18}
D =   {\sum\limits_{-N\leq n<m\leq N } }
 {J_{mn} f_m \left| {f_n } \right|^2 }= - a_\alpha  \left| {f(x,t)} \right|^2 \frac{{\partial ^{\alpha-1} 
f(x,t)}}{{\partial \left| x \right|^{\alpha-1}   }} + \hat J_{\alpha } (0)\left| {f(x,t)} \right|^2f(x,t),
\end{equation}

\begin{equation}\label{eq19}
E ={ \sum\limits_{-N\leq n<m\leq N  }}
{J_{mn}
\left| {f_m } \right|^2 f_n }= - a_\alpha \left| {f(x,t)} \right|^2 \frac{{\partial ^{\alpha-1}   f(x,t)}}{{\partial
\left| x \right|^{\alpha-1}   }}  + \hat{J}_{\alpha  } (0)\left| {f(x,t)} \right|^2f(x,t).
\end{equation}
Next, introducing the following ansatz
\begin{equation}\label{eq21}
f(x,t) = \psi (x,t)\, e^{-i{{2\varepsilon ^2 A t}}/{\hbar }}
\end{equation}
and the parameters
  $U_\alpha   = -F a_\alpha$, 
  $B_\alpha   = Ga_\alpha$, 
 $  D_\alpha   =   {2I -2G\hat J_{\alpha  } (0)}$, equation (\ref{eq10}) now reads
\begin{equation}\label{eq22}
i\hbar \frac{{\partial \psi }}{{\partial t}} + U_\alpha
\frac{{\partial ^{\alpha-1}   \psi }}{{\partial \left| x \right|^{\alpha-1} 
}} + B_\alpha  \psi^2\frac{{\partial ^{\alpha-1}   \psi ^* }}{{\partial
\left| x \right|^{\alpha-1}  }}\ +B_\alpha \left| \psi
\right|^2 \frac{{\partial
^{\alpha-1}   \psi }}{{\partial \left| x \right|^{\alpha-1}   }}  + D_\alpha \left| \psi  \right|^2\psi = 0.
\end{equation}

This is a fractional  cubic nonlinear  Schr\"odinger-like equation that turns out to be the one  governing the dynamics of a ferromagnetic 
Heisenberg spins chain involving long-range interactions in the framework of the approximation described in Appendix B,
i.e. the fields $f$ are slowly varying in space.

However, if the latter
approximation 
is avoided, then
new formulas of the terms $C$ and $E$ are retrieved by the computations
done in Appendix C.
It follows from this analysis that we get a fractional integro-differential cubic nonlinear Schr\"odinger equation given by
\begin{eqnarray}\label{eq23}
i\hbar \frac{{\partial \psi }}{{\partial t}}&+&U_\alpha  \frac{{\partial ^{\alpha-1} 
\psi(x,t)}}{{\partial \left| x \right|^{\alpha-1}   }} +{ B_\alpha\psi^2 \frac{{\partial^{\alpha-1}   \psi^* }}{{\partial \left| x \right|^{\alpha-1}   }}}+{B_\alpha \left| {\psi(x,t)} \right|^2\frac{{\partial ^{\alpha-1} 
\psi(x,t)}}{{\partial \left| x \right|^{\alpha-1}   }}}+{V_\alpha\left| \psi  \right|^2\psi}\nonumber\\
 && -Gb^{\alpha}\int\limits_{ b }^{ + \infty }dy\frac{\left|\psi(x - y,t)\right|^2  \psi(x - y,t)+
\left|\psi(x + y,t)\right|^2\psi(x + y,t)}{{\left|y \right|^{\alpha  } }}\nonumber\\
&& -4b^{\alpha}\int\limits_{ b }^{ + \infty }dy\frac{\left|\psi(x - y,t)\right|^2  \psi(x,t)+\left|\psi(x + y,t)\right|^2\psi(x,t)}{{\left|y \right|^{\alpha  } }}=0,
\end{eqnarray}
with $V_\alpha =- 5G\hat J_{\alpha  }(0)- 2I  $.
Equations~(\ref{eq22}) and~(\ref{eq23}) are the main results of the present paper. 
The approximation done to get the differential Eq.~(\ref{eq22}) corresponds to the excitations with small amplitudes,
while Eq.~(\ref{eq23}) 
can also describe large amplitude excitations.
It would be very interesting, but nontrivial, 
to get their analytical solutions. This task appears \textcolor{black}{not} to be straightforward and in the following we focus on the 
modulational instability of the extended linear solutions of  Eq.~(\ref{eq22}). 


\section{Modulational instability for the continuum medium}

The modulational instability of Eq.~(\ref{eq22}) can be studied using the standard method described for example in \textcolor{black}{Refs.}~\cite{a40,a41}.
Here we are interested in the stability of the homogeneous solution $ \psi(t)=\psi_0\,e^{iD_\alpha\psi_0^{2}t/\hbar}$ \textcolor{black}{for} which the amplitude $\psi_0$ is a real quantity without loss of generality. We remind that we are
in the region of parameters $ 1<\alpha<3 $, with $ \alpha\neq2 $.
 
\begin{figure}[H]
\centering
\includegraphics[width=3.0in,angle=0]{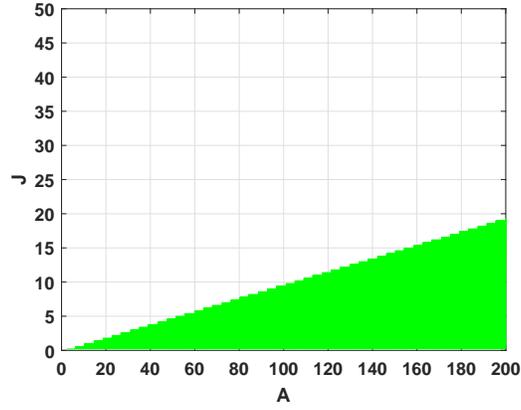}
\caption{\textcolor{black}{Here the green region corresponds the  
      region in the $(A,J$) diagram where the modulational instability could occur.}}
\label{DiagramAJ}
\end{figure}

In the presence of a small perturbation $a(x,t)$ in the system,  one can write
 \begin{equation}\label{eq24}
 \psi(x,t)=(\psi_0+a)e^{iD_\alpha\psi_0^{2} t/\hbar},
\end{equation}
in which $a(x,t)\ll\psi_0$.
Substituting the ansatz~(\ref{eq24}) in Eq.~(\ref{eq22}), we obtain 
\begin{eqnarray}\label{eq25}
i\hbar \frac{{\partial a }}{{\partial t}}&=&-U_\alpha  \frac{{\partial ^{\alpha-1} 
a}}{{\partial \left| x \right|^{\alpha-1}   }} -{ B_\alpha\psi_0 (\psi_0+{2a}) \frac{{\partial^{\alpha-1}   a^* }}{{\partial \left| x \right|^{\alpha-1}   }}}-{B_\alpha \psi_0(\psi_0+{a+a^*})\frac{{\partial ^{\alpha-1} 
a}}{{\partial \left| x \right|^{\alpha-1}   }}}-{D_\alpha\psi_0^{2}(a+a^*)}.
\end{eqnarray}
Next, splitting the perturbation term into real and imaginary parts, $ a(x,t)=u(x,t)+iv(x,t) $, and linearizing Eq.~(\ref{eq25}) with respect to $u$ and $v$, it turns out that we get 
the following system~\textcolor{black}{of equations}
 \begin{eqnarray}\label{eq26}
   \hbar \frac{{\partial u }}{{\partial t}}&=&-U_\alpha  \frac{{\partial ^{\alpha-1} 
v}}{{\partial \left| x \right|^{\alpha-1}   }} , \\
  \hbar \frac{{\partial v }}{{\partial t}}&=&U_\alpha  \frac{{\partial ^{\alpha-1} 
u}}{{\partial \left| x \right|^{\alpha-1}   }} +2{ B_\alpha \psi_0^{2}\frac{{\partial^{\alpha-1}    u }}{{\partial \left| x \right|^{\alpha-1}   }}}+2{D_\alpha\psi_0^{2} u}. 
 \end{eqnarray} 
By introducing the following  Fourier transforms 
\begin{eqnarray}\label{eq27}
   \hat u(k,t)&=&\int\limits_{  - \infty }^{ + \infty } u(x,t) \exp(ixk)dx,\\
  \hat v(k,t)&=&\int\limits_{  - \infty }^{ + \infty } v(x,t) \exp(ixk)dx,
\end{eqnarray}
Equarion~(\ref{eq26}) is converted into a set of ordinary differential equations in the \textcolor{black}{wavevector}  $k$ domain,
\begin{eqnarray}\label{eq28}
   \hbar \frac{{\partial \hat u }}{{\partial t}}&=&U_\alpha \left|k\right|^{\alpha-1} \hat v ,\\
  \hbar \frac{{\partial \hat v }}{{\partial t}}&=&(-U_\alpha-2B_\alpha\psi_0^{2})\left|k\right|^{\alpha-1} \hat u+2{D_\alpha \psi_0^{2}\hat u},
\end{eqnarray}
that can be combined into 
\begin{eqnarray}\label{conbinedequation}
  \hbar \frac{{\partial^2 \hat v }}{{\partial t^2}}&=& 
  \left[2D_\alpha \psi_0^2-(U_\alpha+2B_\alpha\psi_0^2)
  \left|k\right|^{\alpha-1}\right]U_\alpha \left|k\right|^{\alpha-1} \hat v,
\end{eqnarray}
and the same equation for $\hat u$. Consequently, perturbations can grow if and only if the prefactor of the right-hand-side is positive. In such a case, one can thus define
the growth rate of the modulational instability as 
\begin{equation}\label{eq29}
G(k,\alpha)={\sqrt{U_\alpha\left|k\right|^{\alpha-1}\left[2D_\alpha\psi_0^{2}-(U_\alpha+2B_\alpha\psi_0^{2})\left|k\right|^{\alpha-1}\right]}},
\end{equation}
which shows that perturbations with  wavevectors in the range of  $-k_0< k<k_0$ with $k_c=\pm k_0=\pm (2D_\alpha\psi_0^{2}/(U_\alpha+2B_\alpha\psi_0^{2}))^{1/(\alpha-1)}$, are exponentially amplified. 
\textcolor{black}{
  The growth rate, which depends on the parameters $A$ and $J$, the wavevectors $k$ and also the long-range interacting parameter~$\alpha$, is definitely the quantity allowing to characterize the regions of stability/instability of the system. In this specific case, the growth rate can be zero, an imaginary number or either a positive real number. If the growth rate is zero or an imaginary number, then, the system displays stability.  However if the growth rate is a real and positive number, then, the system may  display a modulational instability phenomenon. Therefore, as the modulational instability phenomenon is the main focus of this study, it does occur in specific regions of the values of the parameters.
   Henceforth, Fig.~\ref{DiagramAJ} presents in green the region in the $(A,J)$ plane where the system displays the modulational instability phenomenon.}

\begin{figure}[H]
\centering
\includegraphics[width=3.0in,angle=0]{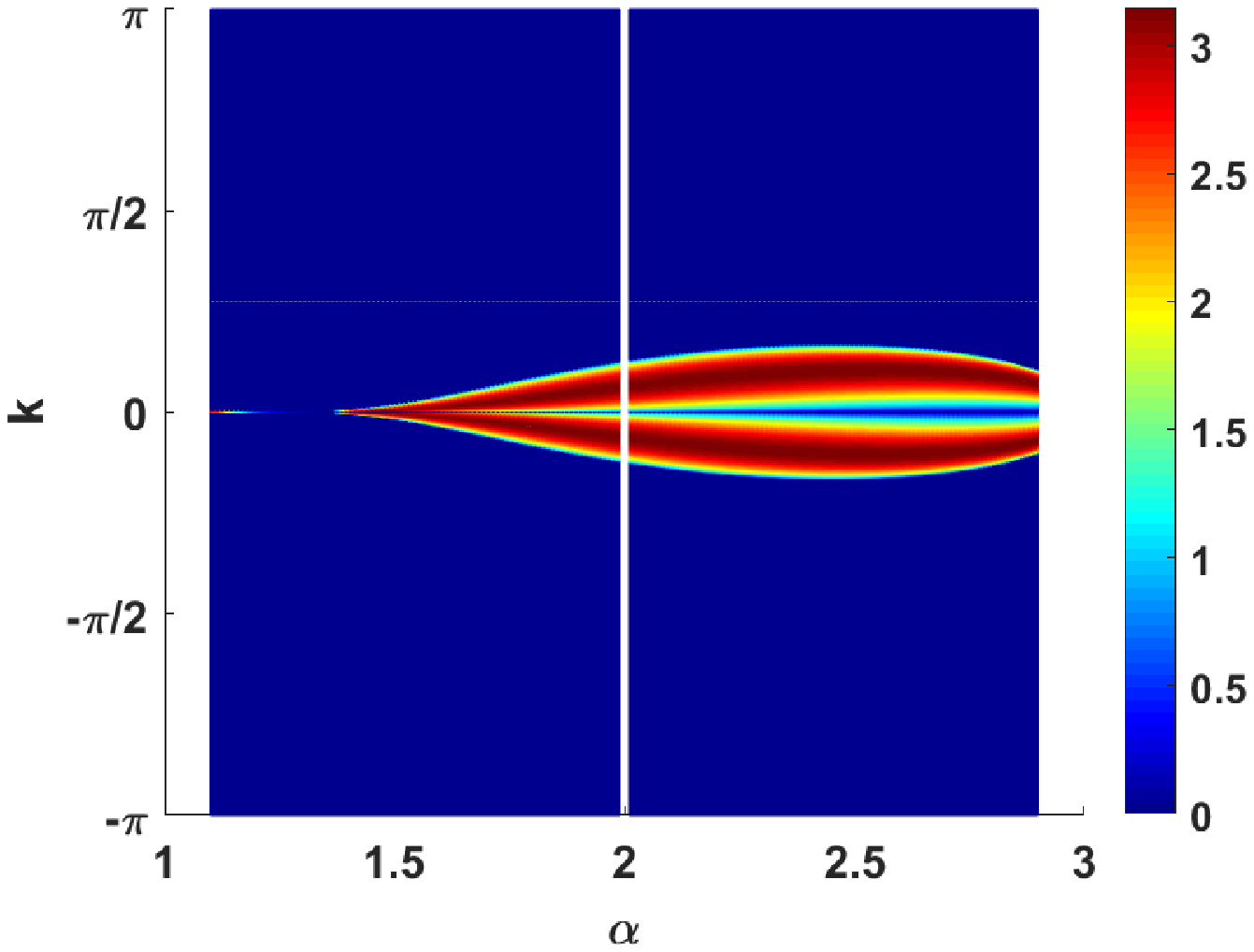}(a)
\includegraphics[width=3.0in,angle=0]{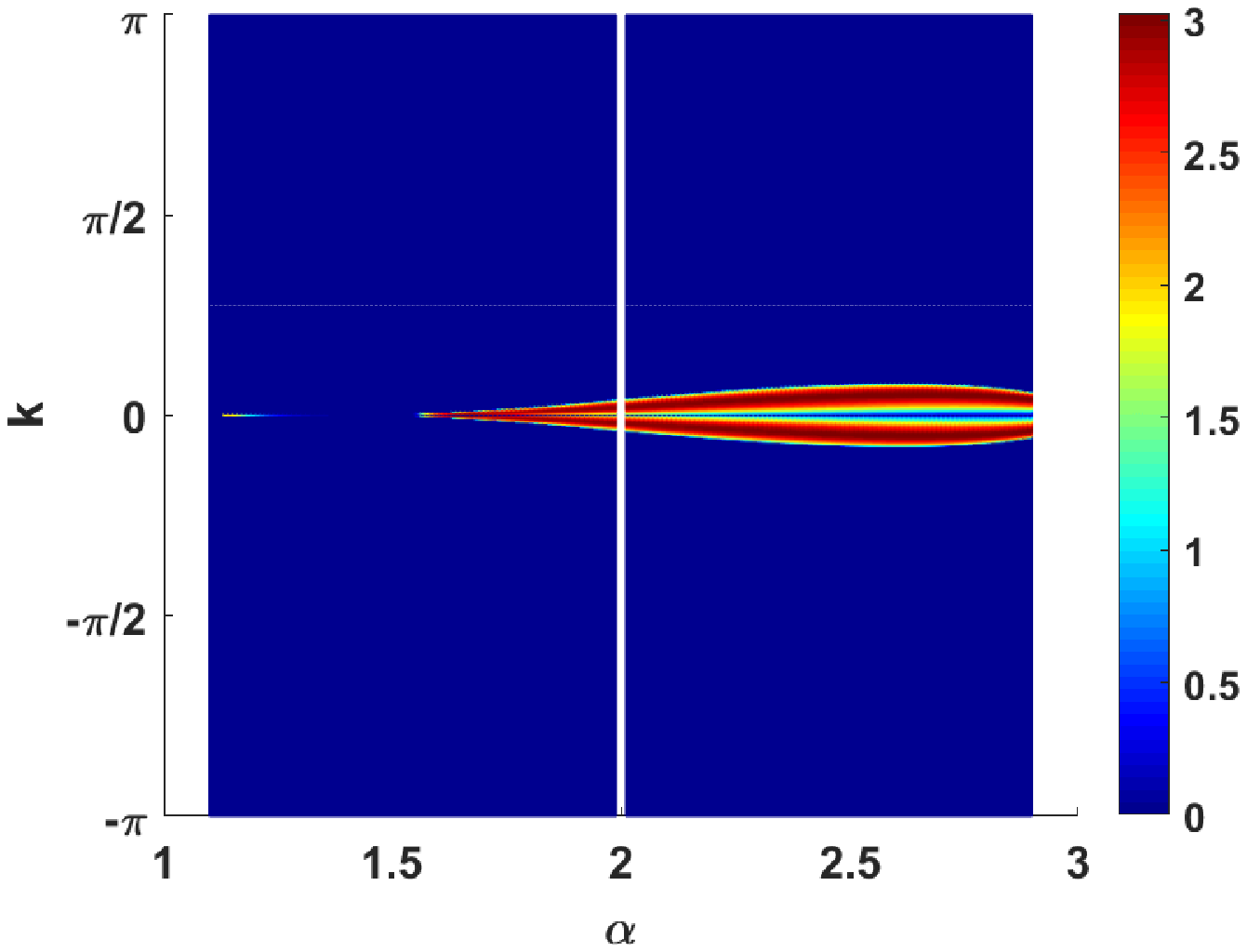}(b)
\caption{\textcolor{black}{The  growth rate of the modulational instability as a function of the long-range interacting exponent $ \alpha $ for different values of $ J=5$ (panel a) and  $J=15$ (b). Both panels are obtained for $A$=160 and $ \psi_0=0.1 $. The blue region corresponds to the stable domain while the multicolor region corresponds to the unstable one.}}
\label{fig2}
\end{figure}

Figure~\ref{fig2} shows the dependence of the modulational instability growth rate on the \textcolor{black}{wavevector}  $k$ and the fractional index~$ \alpha$  for
particular values of the parameter $J$. Here, we realize that the larger is the exchange parameter~$J$, the larger is the stable region.

\begin{figure}[H]
\centering
\includegraphics[width=3.0in,angle=0]{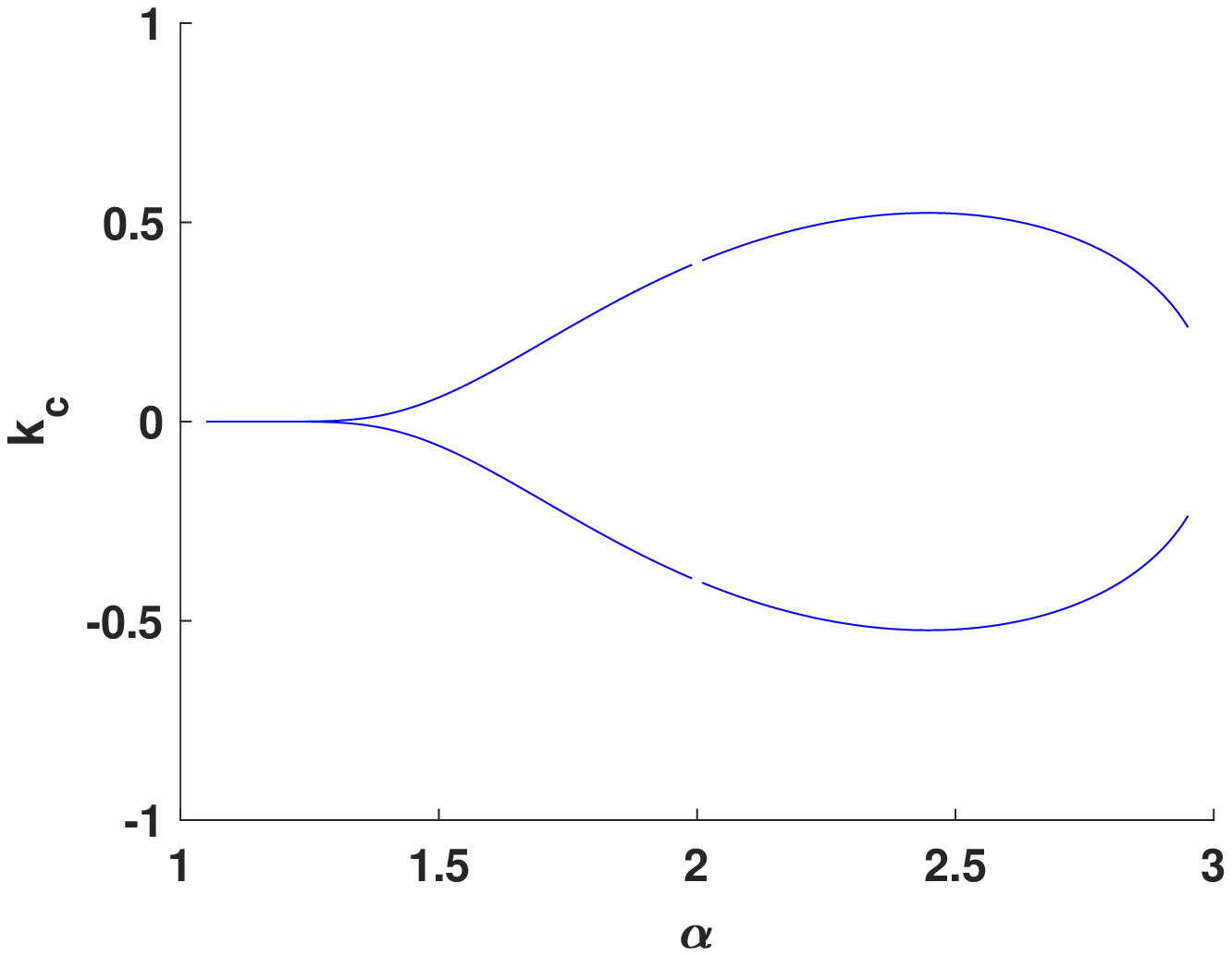}(a)
\includegraphics[width=3.0in,angle=0]{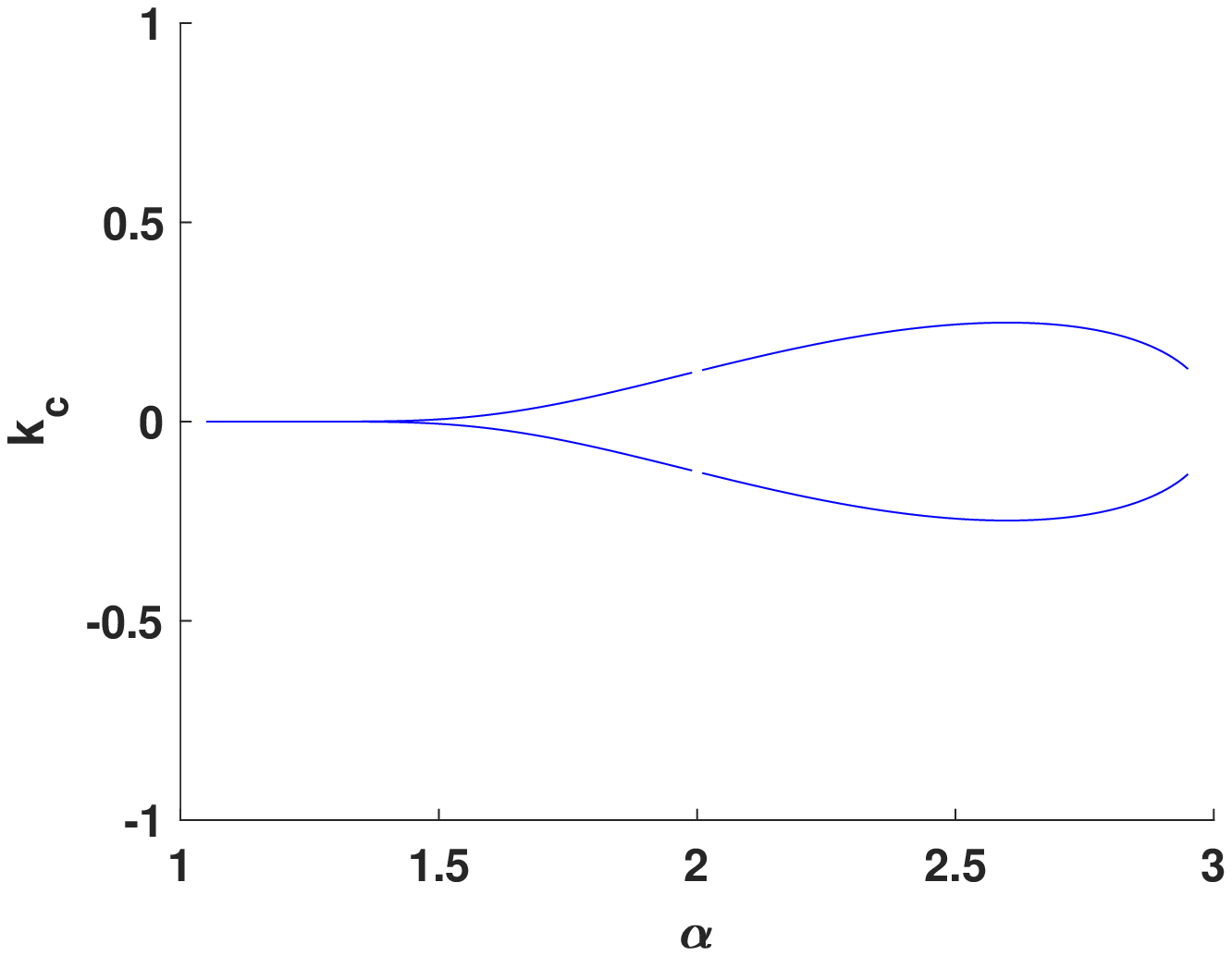}(b)
\caption{\textcolor{black}{The critical frequency  $k_c $ as a function of the exponent $ \alpha $ with $J=5 $ (panel a),  $ J=15 $  (b) and $ \psi_0=0.1 $. Both panels are obtained for $A$=160.} }
\label{fig3}
\end{figure}

{\textcolor{black}Our study is suitable only for non integer values of
  $\alpha$ between $1$ and $3$. While looking at different panels of Fig.~\ref{fig2}  obtained for the same value of the anisotropy parameter $A$, it is realised that as the parameter $J$ increases,the range of values of  $\alpha$ for which the modulational instability occurs, reduces. Figure~\ref{fig3} presents the evolution of the critical wave-vector as a function of the exponent $\alpha$ for the {\textcolor{black}three} cases presented in the previous figure. Similarly, when the parameter $J$ increases as seen in the panels of Fig.~\ref{fig3}, the range of values of the critical wavevactor  $k_c$ reduces.}
  
  \begin{figure}[H]
\centering
\includegraphics[width=3.0in,angle=0]{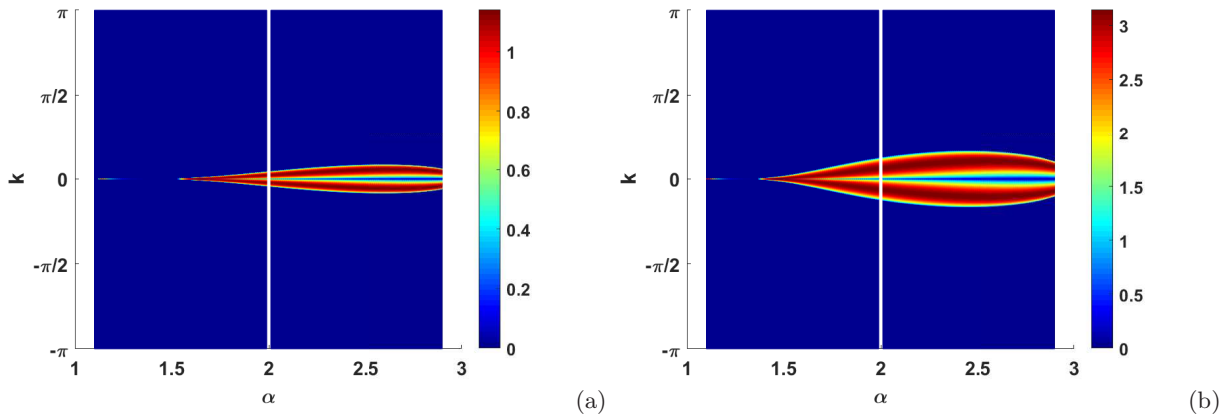}(a)
\includegraphics[width=3.0in,angle=0]{gainrj5A160.eps}(b)
\caption{\textcolor{black}{The  growth rate of the modulational instability as a function of $ \alpha $ for differents values of $ A=60$ (panel a) and $A=160$ (b) while the exchange parameter is fixed to $J=5$ and $ \psi_0=0.1 $. 
The blue region corresponds to the stable domain while the multicolor region corresponds to the unstable one.}}
\label{fig4}
\end{figure}

\begin{figure}[H]
\centering
\includegraphics[width=3.0in,angle=0]{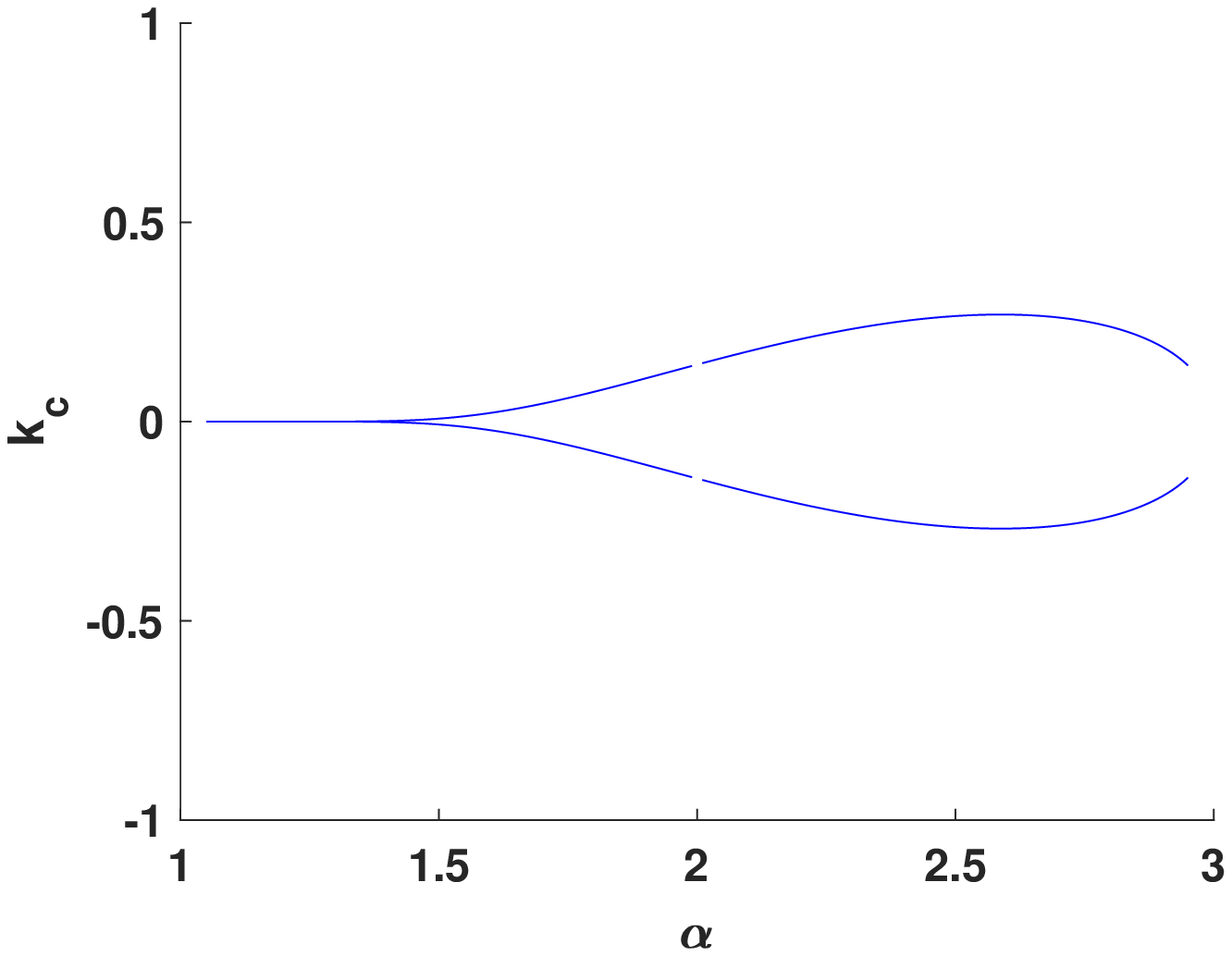}(a)
\includegraphics[width=3.0in,angle=0]{kcj5A160.eps}(b)
\caption{\textcolor{black}{The critical frequency  $k_c $ as a function of the exponent $ \alpha $ with $ A=60 $ (panel a), $ A=160 $  (b) and $ \psi_0=0.1 $. Both panels are obtained for $J$=5.}}
\label{fig5}
\end{figure} 

\textcolor{black}{Figures~\ref{fig4} and~\ref{fig5} show that when the parameter $J$ is fixed and the parameter $A$ varies, the phenomenon is reversed as  compared to what is observed in Fig.~\ref{fig2} and Fig.~\ref{fig3}. This means, as seen in the corresponding panels, that as the anisotropy parameter increases, the instability region increases. 
From this, one sees that if the parameter $J$ tends to favor the stability of the system, the anisotropy parameter at variance induces the instability in such a system. Needless to mention is the fact that while looking at Fig.~\ref{fig2} to Fig.~\ref{fig5}, it is realised that the power-law long-range exponent displays a minimal value below which  the instability does not occur. This minimal value depends also on both the exchange parameter $J$ and the anisotropy parameter $A$. For instance, if the anisotropy parameter $A$ is fixed, the minimal value of $\alpha$ increases with increasing values of the exchange parameter $J$, as seen in panels a) and b) of Fig.~\ref{fig2} and Fig.~\ref{fig3}. The dependence of the minimal value of $\alpha$ on the $A$ and $J$ is also observed when $J$ is fixed and the parameter $A$ varies as seen in panels a) and b) of Fig.~\ref{fig4} and Fig.~\ref{fig5}  with a reversed effect}.

\section{Conclusion} 
We have investigated the effective dynamics of a Heisenberg ferromagnetic spin chain with algebraic long-range couplings using
a semiclassical approximation. We considered a spin chain with long-range power-law interactions 
having a strength proportional to $1/(n-m)^{\alpha}$ in the regime  $1<\alpha<3$ and $ \alpha\neq2 $.  We have used 
the Holstein-Primakoff representation of the spins to derive
a discrete nonlinear cubic Schr\"odinger equation.  We have also shown how one can get, in the continuum limit, on the one hand, a
{\em fractional} cubic nonlinear Schr\"odinger-like equation and, on the other hand, an integro-differential fractional cubic nonlinear 
Schr\"odinger-like equation, corresponding to excitations with small amplitudes and excitations with large amplitudes, respectively. This 
has been achieved after performing the analytical derivatives by firstly using the Riesz derivative of fractional calculus and using, secondly, 
a direct analysis of the Fourier spectrum in the $ k\rightarrow0 $ limit. A remarkable feature of this interaction
is the existence of a transformation that replaces the set of coupled individual bosonic spin equations with a continuum medium equation with 
a fractional spatial derivative of order $\alpha$. Such a transformation is an approximation and it appears in the infrared limit for $ k\rightarrow0 $.  
We also studied the appearance of modulational instability using the obtained fractional equation. The modulational stability has been studied 
for the fractional equation and the stable and unstable regions have been determined. We showed the shifting of the onset of the modulational 
instability region for $ \alpha>1 $ and we studied the dependence of the regions of instability on the parameters of the system. It clearly appears 
that the modulational instability is present only for $\alpha < 3$ and that above $\alpha=3$ only stable regions are present,
thus indicating that the system behaves as a short range system. 
{\textcolor{black}  It is also realized that  the parameter $J$ tends to favorise the stability of the system while the anisotropy parameter $A$ tends to induce the instability in the system.}
As a future work, it would be interesting to compare the results obtained using the effective continuum equation with those of the original
lattice model. In particular it would be rewarding to compare the modulational instability regions obtained with the lattice equation found for
large-$S$ with the findings presented here, since in general the lattice model is expected to have a larger region of instabilities. 
To what extent some unstable phases are missed by the study of the continuum fractional equation is a subject deserving future analytical 
and numerical studies. Finally, we observe that the approach presented here does not apply to $\alpha=2$ and it would be useful to study
in detail the $\alpha \to 2$ limit.

\textcolor{black}{Generalizing this one-dimensional study to 2D and 3D lattices would be very interesting and important.  However, 
 increasing the dimension of the lattice very seriously complicates the derivation of effective equations in the continuum and the search of the solutions. 
 Problems solved in 1D have waited many years, if not decades, to be treated successfully in higher dimensions. However, this would be very helpful and important to check using numerical simulations whether 
 modulational instability thresholds qualitatively depends on the dimension of the lattice.}

\section*{Acknowledgments}

\textcolor{black}{The authors thank CNRS for financial support through the project DyFraLonPo of the ''Dispositif de soutien aux collaborations avec l'Afrique Subsaharienne''.}

\appendix
\section{Properties of the function $J$}
\label{AppendixA}

One thus has to compute the function
\begin{eqnarray}\label{eq30}
  \hat J_{\alpha  } (k) = \sum\limits_{m^{'}  =  - \infty ,m^{'}  \ne 0}^{ + \infty } {\frac{{Je^{ikm^{'} } }}{{\left| {m^{'} } \right|^{\alpha  } }}}  
  &   =&
 \sum\limits_{m^{'}  =  - \infty }^{ - 1} {\frac{{Je^{ikm^{'} } }}{{\left| {m^{'} } \right|^{\alpha  } }}}  + 
        \sum\limits_{m = 1}^{ + \infty } {\frac{{Je^{ikm^{'} } }}{{\left| {m^{'} } \right|^{\alpha  } }}}  \\
 & =& \sum\limits_{m^{'}  = 1}^{ + \infty } {\frac{{Je^{ - ikm^{'} } }}{{\left| {m^{'} } \right|^{\alpha } }}} 
 + \sum\limits_{m = 1}^{ + \infty } {\frac{{Je^{ikm^{'} } }}{{\left| {m^{'} } \right|^{\alpha  } }}}.
\end{eqnarray}
Introducing the  poly-logarithm function defined as
\begin{eqnarray}\label{eq31}
\sum\limits_{m^{'}  =  - \infty ,m^{'}  \ne 0}^{ + \infty } {\frac{{e^{ikm^{'} } }}{{\left| {m^{'} } \right|^{\alpha  } }}}  &=&   {Li_{\alpha  }
(e^{ - ik} ) + Li_{\alpha  } \left( {e^{ik} }
\right)} \\
Li_\alpha  \left( {e^z } \right) &=& \Gamma (1 - \alpha )( -
z)^{\alpha  - 1}  + \sum\limits_{n = 0}^{ + \infty } {\frac{{{\zeta}
\left( {\alpha  - n} \right)}}{{n!}}z^n } ;\qquad\left| z \right| \prec
2\pi
\end{eqnarray}
in which $ {\zeta}$ stands for  the  zeta function defined as
${\zeta} \left( {\alpha  } \right) = \sum\limits_{n = 1}^{ + \infty
} {{1}/{n^\alpha  }}
$
and  
$z = ik\Delta x$, 
it turns out that we get
\begin{eqnarray}\label{eq32}
\hat J_{\alpha  } (k) &=& J\left[ {\Gamma ( 1- \alpha
)\left( {( - ik)^{\alpha-1}    + \left( {ik}
\right)^{\alpha-1}  } \right) + \sum\limits_{n = 0}^{ + \infty }
{\frac{{{\zeta} \left( {\alpha   - n} \right)}}{{n!}}\left( {\left(
{ik} \right)^n  + \left( { - ik} \right)^n }
\right)} } \right]\\
&  =& J\left[ {\Gamma (1 - \alpha )\left|k\right|^{\alpha-1}  \left( {( - i)^{\alpha-1}    + (i)^{\alpha-1}   } \right) 
 + 2\sum\limits_{n = 0}^{ + \infty } {\frac{{{\zeta} \left( {\alpha   - n} \right)}}{{(2n)!}}\left( { - k^2 } \right)^n } } \right]\\
 & =& J\left[ {\Gamma ( 1- \alpha )\left|k\right|^{\alpha-1} (e^{i\frac{\pi }{2}({\alpha-1} ) } 
 + e^{ - i\frac{\pi }{2}({\alpha-1} ) } ) + 2\sum\limits_{n = 0}^{ + \infty } {\frac{{{\zeta} \left( {\alpha   - n} \right)}}{{(2n)!}}\left( { - k^2 } \right)^n } } \right]\\
 & =& 2J\left[ {\Gamma ( 1- \alpha )\left|k\right|^{\alpha-1}  \sin (\frac{\pi }{2}\alpha ) + {\zeta} \left( {\alpha  } \right) 
 + \sum\limits_{n = 1}^{ + \infty } {\frac{{{\zeta} \left( {\alpha   - n} \right)}}{{(2n)!}}\left( { - k^2  } \right)^n } } \right].
\end{eqnarray}
 Denoting $
\hat J_{\alpha  } (0) = 2J{\zeta} \left( {\alpha  } \right)$ and 
$ a_\alpha   = 2J\Gamma (1 - \alpha )\sin \left(\pi \alpha/2 \right)$,
we thus get
\begin{eqnarray}\label{eq33}
 \hat J_{\alpha  } (k) &=& \left[ {a_\alpha  \left|k\right|^{\alpha-1}    +
\hat J_{\alpha  } (0) + 2J\sum\limits_{n = 1}^{ + \infty }
{\frac{{{\zeta} \left( {\alpha   - n} \right)}}{{(2n)!}}\left( { -
k^2 } \right)^n } } \right]\\
\hat J_{\alpha  } (0) - \hat J_{\alpha  } (k) &=&  -
\left[ {a_\alpha  \left|k\right|^{\alpha-1}     + 2J\sum\limits_{n = 1}^{ + \infty }
{\frac{{{\zeta} \left( {\alpha   - n} \right)}}{{(2n)!}}\left( { -
k^2 } \right)^n } } \right].
\end{eqnarray}
In the framework of the continuum approximation \cite{a25}, we get the expression 
\begin{equation}\label{eq60}
\hat J_{\alpha  } (0) - \hat J_{\alpha  } (k) \simeq
- a_\alpha  \left|k\right|^{\alpha-1}  
. \end{equation}
{\textcolor{black} For non-integer $ \alpha>3 $~\cite{a15}, we have
\begin{equation}\label{eq14}
 \hat J_{\alpha  }
(k) \simeq  - \left| k \right|^{2}{{\zeta} \left( {\alpha   - 2} \right)}  + \hat
J_{\alpha  } (0)
\end{equation} }

\section{  Discussion of the different terms of the  discrete nonlinear equation}
\label{AppendixB}

We use the notation
\begin{eqnarray}\label{eq34}
f(x,t) & =\displaystyle \frac{1}{{2\pi }}\int\limits_{ - \infty }^{ + \infty }
 \ \tilde{f}(k,t) {e^{ik\textcolor{black}{ x}} }dk \equiv \mathcal{F}^{-1} \lbrace\tilde{f}(k,t)\rbrace \\
 \tilde{f}(k,t) &= \displaystyle\int\limits_{ - \infty }^{ + \infty }
 \  f(x,t) {e^{-ik\textcolor{black}{ x}} }dk\equiv \mathcal{F} \lbrace f(x,t)\rbrace ,
\end{eqnarray}
where $\mathcal{F} \lbrace f(x,t)\rbrace $ is the Fourier transform of $ f(x,t) $ with respect to $ x $.

In that case, the term $A_1$ defined in Eq.~(\ref{definitionA1}) reads 
\begin{eqnarray}\label{eq35}
A_1  & = \displaystyle\frac{1}{{2\pi }}\int\limits_{ - \infty }^{ + \infty }
{\sum\limits_{m =  - \infty, m \ne n }^{ + \infty } {J_{m,n} } } \ \tilde{f}\left( {e^{ikn}  - e^{ikm} } \right)dk 
\\ & \displaystyle
= \frac{1}{{2\pi }}\int\limits_{ - \infty }^{ + \infty } 
 {\sum\limits_{m = - \infty, m \ne n }^{ + \infty } {\frac{J}{{\left| {m - n} \right|^{\alpha  } }}\tilde{f}\left( {k,t} \right)\left( {e^{ikn}  - e^{ikm} } \right)} dk}.
\end{eqnarray}
Setting then
$m^{'}  = m - n$, 
we get the following expression
\begin{eqnarray}\label{eq36}
A_1 & =& \frac{1}{{2\pi }}\int\limits_{ - \infty }^{ + \infty }
{\sum\limits_{m^{'}  =  - \infty ,m^{'}  \ne 0}^{ + \infty }
{\frac{J}{{\left| {m^{'} } \right|^{\alpha  } }}\tilde{f}\left(
{k,t} \right)\left( {e^{ikn}  - e^{ik(m^{'}  + n)}
} \right)} dk}\\
 & =& \frac{1}{{2\pi }}\left(\int\limits_{ - \infty }^{ + \infty } {\sum\limits_{m^{'}  =  - \infty ,m^{'}  \ne 0}^{ + \infty } {\frac{J}{{\left| {m^{'} } \right|^{\alpha  } }}\tilde{f}(k,t)e^{ikn} } dk}  - \int\limits_{ - \infty }^{ + \infty } {\sum\limits_{m^{'}  =  - \infty }^{ + \infty } {\frac{{Je^{ikm^{'} } }}{{\left| {m^{'} } \right|^{\alpha   } }}\tilde{f}(k,t)e^{ikn} } dk}\right)\\
  & =& \frac{1}{{2\pi }}\left[ {\int\limits_{ - \infty }^{ + \infty } {\hat J_{\alpha  } \left( 0 \right)\tilde{f}(k,t)e^{ikn} dk}  - \int\limits_{ - \infty }^{ + \infty } {\hat J_{\alpha  } \left( {k} \right)\tilde{f}(k,t)e^{ikn} } dk)} \right]\\
 & =&\frac{1}{{2\pi }}\int\limits_{ - \infty }^{ + \infty } {(\hat J_\alpha  (0) - \hat J_\alpha  (k))\tilde{f}(k,t)e^{ikn} dk},
\end{eqnarray}
where we set  
$ x=k\Delta x$ and 
 $\hat J_{\alpha  } ( x) = \sum\limits_{m^{'}  =  - \infty ,m^{'}  \ne 0}^{ + \infty }J{{{ e^{im^{'} x} }}/{{\left| {m^{'} } \right|^{\alpha  } }}}  
$. 
 It turns out that we get
\begin{equation}\label{eq38}
A_1 =  - \frac{{a_{\alpha}  }}{{2\pi }}\int\limits_{ - \infty }^{ +\infty } {k^{\alpha-1}  \hat f(k,t)e^{ikn} dk}.
\end{equation}
 Since the Fourier transform involving the absolute value of momentum $\left|k\right|^\alpha$ is expressed by Riesz derivative in the real space as
\begin{eqnarray}\label{eq39}
 \frac{1}{{2\pi }}\int\limits_{ - \infty }^{ + \infty } {\left|k\right|^\alpha  \tilde{f}(k,t)e^{ikn} dk}  \simeq  - \frac{{\partial ^\alpha  f(x,t)}}{{\partial \left| x \right|^\alpha  }} 
\end{eqnarray}
with $x = n\Delta x$,
 the term $A_1$ finally reads
\begin{equation}\label{eq40}
 A_1 = a_\alpha  \frac{{\partial ^{\alpha-1}
f(x,t)}}{{\partial \left| x \right|^{\alpha-1} }}.
\end{equation}
The term $C$ can be split as
\begin{equation}\label{eq41}
C = \underbrace{\sum\limits_{m =  - \infty ,m \ne n}^{ n-1 } {\frac{J}
{{\left| {m - n} \right|^\alpha  }}\left| {f_m } \right|^2f_m  }}_{C_1}  + 
\underbrace{\sum\limits_{m = n+1,m \ne n}^{ + \infty } {\frac{J}
{{\left| {m - n} \right|^\alpha  }} \left| {f_m } \right|^2 f_m} }_{C_2}
\end{equation}
that  can be analytically computed  separately.  Then, we get
\begin{equation}\label{eq42}
 C_1= \left( {\frac{1}{{2\pi }}} \right)^3 \int\limits_{ - \infty }^{ + \infty } 
 {\sum\limits_{m =  - \infty ,m \ne n}^{ n-1 } {\frac{{Je^{ikm} }}{{\left| {m-n } \right|^{\alpha  } }}} \tilde{f}(k,t) dk\int\limits_{ - \infty }^{ + \infty } {\tilde{f}^ *  (k^{'} ,t)e^{ - ik^{'} (m)} dk^{'} \int\limits_{ - \infty }^{ + \infty } {\tilde{f}(k^{''} ,t)e^{ik^{''} (m)} dk^{''} } } }
 \end{equation}
  and next  setting
$m - n = m^{'} $,  it turns out that
\begin{eqnarray}\label{eq43}
  C_1 &=& \left( {\frac{1}{{2\pi }}} \right)^3 \int\limits_{ - \infty }^{ + \infty }\!\! {\sum\limits_{m^{'} =  - \infty ,m^{'} \ne 0}^{ -1} \!\!\!\!\!\!{\frac{{Je^{ikm^{'} } }}{{\left| {m^{'} } \right|^{\alpha  } }}} \tilde{f}(k,t)e^{ikn} dk \!\! \int\limits_{ - \infty }^{ + \infty } {\tilde{f}^ *  (k^{'} ,t)e^{ - ik^{'} (m^{'}  + n)} dk^{'} \!\! \int\limits_{ - \infty }^{ + \infty } {\tilde{f}(k^{''} ,t)e^{ik^{''} (m^{'}  + n)} dk^{''} } } }\\
  & =& \left( {\frac{1}{{2\pi }}} \right)^3 \int {\int {\int_{ - \infty }^{ + \infty } {\sum\limits_{m^{'} =  1 ,m^{'} \ne 0}^{ + \infty } \!\!\!\!{\frac{{Je^{-i(k - k^{'}  + k^{''} )m^{'} } }}{{\left| {m^{'} } \right|^{\alpha  } }}} \tilde{f}(k,t)e^{ikn} \tilde{f}^ *  (k^{'} ,t)e^{ - ik^{'} n} \tilde{f}(k^{''} ,t)e^{ik^{''} n} dkdk^{'} dk^{''} } } }\\
   & =& \left( {\frac{1}{{2\pi }}} \right)^3 \int\limits_{ - \infty }^{ + \infty }\!\! {\sum\limits_{m^{'} = 1 ,m^{'} \ne 0}^{ + \infty } \!\!\!\!\!\!{\frac{{Je^{-ikm^{'} \Delta x} }}{{\left| {m^{'} } \right|^{\alpha  } }}} \tilde{f}(k,t)e^{ikn} dk \!\!\!\! \int\limits_{ - \infty }^{ + \infty } {\tilde{f}^ *  (k^{'} ,t)e^{ - ik^{'} (-m^{'}  + n)} dk^{'} \!\!\!\!\int\limits_{ - \infty }^{ + \infty } {\tilde{f}(k^{''} ,t)e^{ik^{''} (-m^{'}  +} dk^{''} } } }\\
   \label{eq44}
C_1 & =& \left( {\frac{1}{{2\pi }}} \right) \int\limits_{ - \infty }^{ + \infty } {\sum\limits_{m^{'} = 1 ,m^{'} \ne 0}^{ + \infty } {\frac{{Je^{-ikm^{'} } }}{{\left| {m^{'} } \right|^{\alpha  } }}} \tilde{f}(k,t)e^{ikn} dk}f_{n - m^{'}}^* f_{n - m^{'}}.
\end{eqnarray}

Before moving to the continuum limit, we should remind that there are terms with  sums of  type $\sum\limits_{m^{'}} {f_{n \pm m^{'}}} $.  In this respect, we can proceed to the approximation based on the assumption that in these sums, the fields ${f_{n \pm m^{'}} }$'s are slowly varying in space. Therefore,  such terms as ${f_{n \pm m^{'}} }$  can be brought outside the sums over $m^{'}$~\cite{a7}.
In this case, we can now write
\begin{equation}\label{eq45}
C_1= f_{n - m^{'}}^* f_{n - m^{'}}\left( {\frac{1}{{2\pi }}} \right) \int\limits_{ - \infty }^{ + \infty } {\sum\limits_{m^{'} = 1 ,m^{'} \ne 0}^{ + \infty } {\frac{{Je^{-ikm^{'} } }}{{\left| {m^{'} } \right|^{\alpha  } }}} \tilde{f}(k,t)e^{ikn} dk} .
\end{equation}
Following the same later steps, we get
\begin{eqnarray}\label{eq46}
C_2  
  &=& \left( {\frac{1}{{2\pi }}} \right)^3 \int\limits_{ - \infty }^{ + \infty } {\sum\limits_{m =  n+1 ,m^ \ne n}^{ + \infty } {\frac{{Je^{ikm} }}{{\left| {m-n } \right|^{\alpha  } }}} \tilde{f}(k,t) dk\int\limits_{ - \infty }^{ + \infty } {\tilde{f}^ *  (k^{'} ,t)e^{ - ik^{'} (m)} dk^{'} \int\limits_{ - \infty }^{ + \infty } {\tilde{f}(k^{''} ,t)e^{ik^{''} (m)} dk^{''} } } }\\
  &=& \left( {\frac{1}{{2\pi }}} \right)^3 \int\limits_{ - \infty }^{ + \infty } {\sum\limits_{m^{'} =  1 ,m^{'} \ne 0}^{ + \infty} {\frac{{Je^{ikm^{'} } }}{{\left| {m^{'} } \right|^{\alpha  } }}} \tilde{f}(k,t)e^{ikn} dk\int\limits_{ - \infty }^{ + \infty } {\tilde{f}^ *  (k^{'} ,t)e^{ - ik^{'} (m^{'}  + n)} dk^{'} \int\limits_{ - \infty }^{ + \infty } {\tilde{f}(k^{''} ,t)e^{ik^{''} (m^{'}  + n)} dk^{''} } } },
\end{eqnarray}
that can be rewritten as
\begin{equation}\label{eq47}
C_2= \left( {\frac{1}{{2\pi }}} \right) \int\limits_{ - \infty }^{ + \infty } {\sum\limits_{m^{'} = 1 ,m^{'} \ne 0}^{ + \infty } {\frac{{Je^{ikm^{'} } }}{{\left| {m^{'} } \right|^{\alpha  } }}} \tilde{f}(k,t)e^{ikn} dk}f_{n + m^{'}}^* f_{n + m^{'}} .
\end{equation}
Then, if we consider here also  the same  approximation used above to move from Eq.~(\ref{eq44})  to Eq.~(\ref{eq45}), we get
 \begin{equation}\label{eq48}
C_2= f_{n + m^{'}}^* f_{n + m^{'}}\left( {\frac{1}{{2\pi }}} \right) \int\limits_{ - \infty }^{ + \infty } {\sum\limits_{m^{'} = 1 ,m^{'} \ne 0}^{ + \infty } {\frac{{Je^{-ikm^{'} } }}{{\left| {m^{'} } \right|^{\alpha  } }}} \tilde{f}(k,t)e^{ikn} dk} .
\end{equation}
It then follows that $C$ is given by
\begin{equation}\label{eq49}
C= f_{n - m^{'}}^* f_{n - m^{'}}\left( {\frac{1}{{2\pi }}} \right) \int\limits_{ - \infty }^{ + \infty } {\sum\limits_{m^{'} = 1 ,m^{'} \ne 0}^{ + \infty } {\frac{{Je^{-ikm^{'} } }}{{\left| {m^{'} } \right|^{\alpha  } }}} \tilde{f}(k,t)e^{ikn} dk}+ \\
f_{n + m^{'}}^* f_{n + m^{'}}\left( {\frac{1}{{2\pi }}} \right) \int\limits_{ - \infty }^{ + \infty } {\sum\limits_{m^{'} = 1 ,m^{'} \ne 0}^{ + \infty } {\frac{{Je^{ikm^{'} } }}{{\left| {m^{'} } \right|^{\alpha  } }}} \tilde{f}(k,t)e^{ikn} dk} 
\end{equation}
As 
$n \pm m ^{'}\simeq n$, 
we get
\begin{equation}\label{eq50}
 C= f_{n }^* f_{n }\left( {\frac{1}{{2\pi }}} \right) \int\limits_{ - \infty }^{ + \infty } {\sum\limits_{m^{'} = 1 ,m^{'} \ne 0}^{ + \infty } {\frac{{Je^{-ikm^{'} }}}{{\left| {m^{'} } \right|^{\alpha  } }}}\tilde{f}(k,t)e^{ikn} dk} +f_{n }^* f_{n }\left( {\frac{1}{{2\pi }}} \right) \int\limits_{ - \infty }^{ + \infty } {\sum\limits_{m^{'} = 1 ,m^{'} \ne 0}^{ + \infty } {\frac{{Je^{ikm^{'} } }}{{\left| {m^{'} } \right|^{\alpha  } }}}\tilde{f}(k,t)e^{ikn} dk}
\end{equation}
which leads us to
\begin{eqnarray}\label{eq51}
 C&=& f_{n }^* f_{n }\left( {\frac{1}{{2\pi }}} \right) \int\limits_{ - \infty }^{ + \infty } {\sum\limits_{m^{'} = 1 ,m^{'} \ne 0}^{ + \infty } {\frac{{Je^{-ikm^{'} }}+{Je^{ikm^{'} \Delta x}}}{{\left| {m^{'} } \right|^{\alpha  } }}} \tilde{f}(k,t)e^{ikn} dk} \\
 \label{eq53}
 &=& f_{n }^* f_{n }\left( {\frac{1}{{2\pi }}} \right) \int\limits_{ - \infty }^{ + \infty }{\hat J_{\alpha  } ( {k} )  \tilde{f}(k,t)e^{ikn} dk},
\end{eqnarray}
as\begin{equation}\label{eq52}
\hat J_{\alpha  } \left( {k} \right) ={\sum\limits_{m^{'} = 1 ,m^{'} \ne 0}^{ + \infty } {\frac{{Je^{-ikm^{'} }}+{Je^{ikm^{'} }}}{{\left| {m^{'} } \right|^{\alpha  } }}}}.
\end{equation}
Here we consider the continuum approximation done in Eq.~(\ref{eq13}) to get 
\begin{eqnarray}\label{eq54}
 C &\simeq & f_{n }^* f_{n }\left( {\frac{1}{{2\pi }}} \right) \int\limits_{ - \infty }^{ + \infty } { \left(a_\alpha \left|k\right|^{\alpha-1}   
+ \hat J_{\alpha  } (0)\right) \hat f(k,t)e^{ikn\Delta x} dk} \\
  & =&f_{n }^* f_{n }\left( {\frac{1}{{2\pi }}} \right) \int\limits_{ - \infty }^{ + \infty } { a_\alpha \left|k\right|^{\alpha-1} }  \tilde{f}(k,t)e^{ikn} dk +f_{n }^* f_{n }\left( {\frac{1}{{2\pi }}} \right) \int\limits_{ - \infty }^{ + \infty } {{\hat J_{\alpha  } (0)} \tilde{f}(k,t)e^{ikn} dk}.
\end{eqnarray}
Considering that 
$f_n  = f(x = n\Delta x,t)$, the  continuum approximation using the Riesz derivative gives by Eq.~(\ref{continuumalpha9})  leads us to
\begin{equation}\label{eq55}
C =  - a_\alpha  \frac{{\partial ^{\alpha-1}   f(x,t)}}{{\partial
\left| x \right|^{\alpha-1} }}\left| {f(x,t)} \right|^2  + \hat
J_{\alpha  } (0)\left| {f(x,t)} \right|^2f(x,t).
\end{equation}

Then splitting $ E $ as follows,
\begin{equation}\label{eq56}
E  = \sum\limits_{m =  - \infty ,m \ne n}^{ + \infty } {J_{mn} \left| {f_m } \right|^2 f_n }
=\underbrace{\sum\limits_{m =  - \infty ,m \ne n}^{n-1}  {J_{mn}
\left| {f_m } \right|^2 f_n }}_{E_1} + \underbrace{\sum\limits_{m = n+1,m \ne n}^{ + \infty } {J_{mn}
\left| {f_m } \right|^2 f_n }}_{E_2},
\end{equation}
one has to calculate each term. 
\begin{eqnarray}\label{eq57}
E_1 &=&\sum\limits_{m =  - \infty ,m \ne n}^{n-1}  {J_{mn}
\left| {f_m } \right|^2 f_n } \\
  &=& \left( {\frac{1}{{2\pi }}} \right)^3 \int\limits_{ - \infty }^{ + \infty }{\sum\limits_{m =  - \infty ,m \ne n}^{n-1}    {\frac{J}{{\left| {m - n} \right|^{\alpha  } }}} \tilde{f}(k,t)e^{ikm} dk\int\limits_{ - \infty }^{ + \infty } {\tilde{f}^* (k^{'} ,t)e^{ - ik^{'} m} dk^{'} \int\limits_{ - \infty }^{ + \infty } {\tilde{f}(k^{''} ,t)e^{ik^{''} n} dk^{''} } } }.
\end{eqnarray}
now, setting 
$m - n = m^{'} $, we get 
\begin{eqnarray}\label{eq58}
E_1   &=& \left( {\frac{1}{{2\pi }}} \right)^3 \int\limits_{ - \infty }^{ + \infty } {\sum\limits_{m ^{'}=  - \infty ,m^{'} \ne 0}^{ -1} {\frac{{Je^{ikm^{'} } }}{{\left| {m^{'} } \right|^{\alpha  } }}}\tilde{f}(k,t)e^{ikn} dk\int\limits_{ - \infty }^{ + \infty } {\tilde{f}(k^{'} ,t)e^{ - ik^{'} (m^{'}  + n)} dk^{'} \int\limits_{ - \infty }^{ + \infty } {\tilde{f}(k^{''} ,t)e^{ik^{''} n} dk^{''} } } }\\
 &= &\left( {\frac{1}{{2\pi }}} \right)^3 \int\limits_{ - \infty }^{ + \infty } {\sum\limits_{m ^{'}=  - \infty ,m^{'} \ne 0}^{ -1} {\frac{{Je^{i(k-k^{'})m^{'} } }}{{\left| {m^{'} } \right|^{\alpha  } }}} \tilde{f}(k,t)e^{ikn} dk\int\limits_{ - \infty }^{ + \infty } {\tilde{f}(k^{'} ,t)e^{ - ik^{'}n\Delta x} dk^{'} \int\limits_{ - \infty }^{ + \infty } {\tilde{f}(k^{''} ,t)e^{ik^{''} n} dk^{''} } } }\\
& =& \left( {\frac{1}{{2\pi }}} \right)^3 \int\limits_{ - \infty }^{ + \infty } {\sum\limits_{m ^{'}=  1 ,m^{'} \ne 0}^{+\infty } {\frac{{Je^{-i(k-k^{'})m^{'} } }}{{\left| {m^{'} } \right|^{\alpha  } }}} \tilde{f}(k,t)e^{ikn} dk\int\limits_{ - \infty }^{ + \infty } {\tilde{f}(k^{'} ,t)e^{ - ik^{'}n\Delta x} dk^{'} \int\limits_{ - \infty }^{ + \infty } {\tilde{f}(k^{''} ,t)e^{ik^{''} n} dk^{''} } } }\\
&  = &\left( {\frac{1}{{2\pi }}} \right)^3 \int\limits_{ - \infty }^{ + \infty } {\sum\limits_{m ^{'}=  1 ,m^{'} \ne 0}^{+\infty } {\frac{{Je^{-ikm^{'} } }}{{\left| {m^{'} } \right|^{\alpha  } }}} \tilde{f}(k,t)e^{ikn} dk\int\limits_{ - \infty }^{ + \infty } {\tilde{f}(k^{'} ,t)e^{ - ik^{'}(n-m^{'})} dk^{'} \int\limits_{ - \infty }^{ + \infty } {\tilde{f}(k^{''} ,t)e^{ik^{''} n} dk^{''} } } }\\
& = &\left( {\frac{1}{{2\pi }}} \right) \int\limits_{ - \infty }^{ + \infty } {\sum\limits_{m ^{'}=  1 ,m^{'} \ne 0}^{+\infty } {\frac{{Je^{-ikm^{'} } }}{{\left| {m^{'} } \right|^{\alpha  } }}} \tilde{f}(k,t)e^{ikn} dk }f_{n - m^{'}}^* f_{n }.
 \end{eqnarray}
 Then, proceeding to the same  approximation used  to move from Eq.~(\ref{eq44})  to Eq.~(\ref{eq45}), we  get 
  \begin{equation}\label{eq59} 
 E_1= \left( {\frac{1}{{2\pi }}} \right)f_{n - m^{'}}^* f_{n } \int\limits_{ - \infty }^{ + \infty } {\sum\limits_{m ^{'}=  1 ,m^{'} \ne 0}^{+\infty } {\frac{{Je^{-ikm^{'} } }}{{\left| {m^{'} } \right|^{\alpha  } }}} \tilde{f}(k,t)e^{ikn} dk } .
 \end{equation}
Through the same approximation and very similar steps, one derives 
\begin{equation}\label{eq60_bis}
 E_2= \left( {\frac{1}{{2\pi }}} \right)f_{n + m^{'}}^* f_{n } \int\limits_{ - \infty }^{ + \infty } {\sum\limits_{m ^{'}=  1 ,m^{'} \ne 0}^{+\infty } {\frac{{Je^{ikm^{'} } }}{{\left| {m^{'} } \right|^{\alpha  } }}} \tilde{f}(k,t)e^{ikn} dk } 
 \end{equation}
that leads to  \begin{eqnarray}\label{eq61}
 E &=&  \left( {\frac{1}{{2\pi }}} \right)f_{n - m^{'}}^* f_{n } \int\limits_{ - \infty }^{ + \infty } {\sum\limits_{m ^{'}=  1 ,m^{'} \ne 0}^{+\infty } {\frac{{Je^{-ikm^{'} } }}{{\left| {m^{'} } \right|^{\alpha  } }}} \tilde{f}(k,t)e^{ikn\Delta x} dk } \nonumber \\
&&+ \left( {\frac{1}{{2\pi }}} \right)f_{n + m^{'}}^* f_{n } \int\limits_{ - \infty }^{ + \infty } {\sum\limits_{m ^{'}=  1 ,m^{'} \ne 0}^{+\infty } {\frac{{Je^{ikm^{'} } }}{{\left| {m^{'} } \right|^{\alpha  } }}} \tilde{f}(k,t)e^{ikn} dk } \\
 &= & \left( {\frac{1}{{2\pi }}} \right)f_{n }^* f_{n } \int\limits_{ - \infty }^{ + \infty } {\sum\limits_{m ^{'}=  1 ,m^{'} \ne 0}^{+\infty } {\frac{{Je^{-ikm^{'} } }+{Je^{ikm^{'} \Delta x} }}{{\left| {m^{'} } \right|^{\alpha  } }}} \tilde{f}(k,t)e^{ikn} dk }  \\
  & =&  \left( {\frac{1}{{2\pi }}} \right)f_{n }^* f_{n } \int\limits_{ - \infty }^{ + \infty } {\hat J_{\alpha  } ( {k} )  \tilde{f}(k,t)e^{ikn\Delta x} dk }\\
 &=&f_{n }^* f_{n }\left( {\frac{1}{{2\pi }}} \right) \int\limits_{ - \infty }^{ + \infty } { a_\alpha (k)^{\alpha-1} }  \tilde{f}(k,t)e^{ikn} dk +f_{n }^* f_{n }\left( {\frac{1}{{2\pi }}} \right) \int\limits_{ - \infty }^{ + \infty } {{\hat J_{\alpha  } (0)} \tilde{f}(k,t)e^{ikn} dk}.
  \end{eqnarray}
   When we replace all these terms in the equation of motion (\ref{eq10}), we get
\begin{eqnarray}\label{eq62}
i\hbar \dot f &=& F a_\alpha  \frac{{\partial ^{\alpha-1}   f(x,t)}}{{\partial \left| x \right|^{\alpha-1}   }}-  Tf + 2I\left| f \right|^2 f\nonumber \\
&&+ G\left[ { - a_\alpha  \frac{{\partial
^{\alpha-1}   f^* }}{{\partial \left| x \right|^{\alpha-1}  }}\
+ \hat J_{\alpha  } (0)\left| f \right|^2 f - a_\alpha \left| f \right|^2 \frac{{\partial ^{\alpha-1}   f}}{{\partial \left| x \right|^{\alpha-1}   }}  
+ \hat J_{\alpha  } (0)\left| f \right|^2f   + 4\hat J_{\alpha  } (0)\left| f \right|^2f  - 4\hat J_{\alpha  } (0)\left| f \right|^2f} \right].
\end{eqnarray}

\section{The fractional integro-differential nonlinear equation}
\label{AppendixC}

One has
\begin{eqnarray}\label{eq63}
C &=&\left( {\frac{1}{{2\pi }}} \right) \int\limits_{ - \infty }^{ + \infty } {\sum\limits_{m^{'} = 1 ,m^{'} \ne 0}^{ + \infty } {\frac{{Je^{-ikm^{'} } }}{{\left| {m^{'} } \right|^{\alpha  } }}} \tilde{f}(k,t)e^{ikn} dk}f_{n - m^{'}}^* f_{n - m^{'}}+\left( {\frac{1}{{2\pi }}} \right) \int\limits_{ - \infty }^{ + \infty } {\sum\limits_{m^{'} = 1 ,m^{'} \ne 0}^{ + \infty } {\frac{{Je^{ikm^{'} } }}{{\left| {m^{'} } \right|^{\alpha  } }}} \tilde{f}(k,t)e^{ikn} dk}f_{n + m^{'}}^* f_{n + m^{'}}\nonumber \\
 &=&\left( {\frac{1}{{2\pi }}} \right)\sum\limits_{m^{'} = 1 ,m^{'} \ne 0}^{ + \infty }f_{n - m^{'}}^* f_{n - m^{'}}  \int\limits_{ - \infty }^{ + \infty } {{\frac{{Je^{ik(n-m^{'}) } }}{{\left| {m^{'} } \right|^{\alpha  } }}} \tilde{f}(k,t) dk}+\left( {\frac{1}{{2\pi }}} \right)\sum\limits_{m^{'} = 1 ,m^{'} \ne 0}^{ + \infty }f_{n + m^{'}}^* f_{n + m^{'}} \int\limits_{ - \infty }^{ + \infty } { {\frac{{Je^{ik(m^{'}+n) } }}{{\left| {m^{'} } \right|^{\alpha  } }}} \tilde{f}(k,t)dk}.
\end{eqnarray}
In the continuum limit $ \Delta x\rightarrow0$, by replacing $n$ and $m$ respectively by $ {x}/{\Delta x} $ and $ {y}/{\Delta x} $, we  get
\begin{eqnarray}\label{eq64}
 C &=&\left( {\frac{1}{{2\pi }}} \right)\int\limits_{ b }^{ + \infty }dy\left|f(x - y,t)\right|^2  \int\limits_{ - \infty }^{ + \infty } {{\frac{{Je^{i\frac{k}{b}(x-y) } }}{{\left| {\frac{y}{b} } \right|^{\alpha  } }}} \tilde{f}(k,t) dk}+\left( {\frac{1}{{2\pi }}} \right)\int\limits_{ b }^{ + \infty }dy\left|f(x + y,t)\right|^2  \int\limits_{ - \infty }^{ + \infty } {{\frac{{Je^{i\frac{k}{b}(x+y) } }}{{\left| {\frac{y}{b} } \right|^{\alpha  } }}} \tilde{f}(k,t) dk}\nonumber \\
 &=&\left( {\frac{1}{{2\pi }}} \right)\int\limits_{ b }^{ + \infty }dy\left|f(x - y,t)\right|^2 b^{\alpha} \int\limits_{ - \infty }^{ + \infty } {{\frac{{Je^{ip(x-y) } }}{{\left| y  \right|^{\alpha  } }}} \tilde{f}(k,t) dk}+\left( {\frac{1}{{2\pi }}} \right)\int\limits_{ b }^{ + \infty }dy\left|f(x + y,t)\right|^2 b^{\alpha} \int\limits_{ - \infty }^{ + \infty } {{\frac{{Je^{ip(x+y) } }}{{\left| y \right|^{\alpha  } }}} \tilde{f}(k,t) dk}\nonumber \\
  &=&b^{\alpha}\int\limits_{ b }^{ + \infty }dy\frac{\left|f(x - y,t)\right|^2  f(x - y,t)+\left|f(x + y,t)\right|^2f(x + y,t)}{{\left|y \right|^{\alpha  } }}.
\end{eqnarray}
where $ b=\Delta x $ and $p= {k}/{b} $.
In the same spirit of analytical computations, the term $E$ will be given by
\begin{eqnarray}\label{eq65}
E &=&\left( {\frac{1}{{2\pi }}} \right)\sum\limits_{m ^{'}=  1 ,m^{'} \ne 0}^{+\infty }f_{n - m^{'}}^* f_{n } \int\limits_{ - \infty }^{ + \infty } { {\frac{{Je^{ik(n-m^{'}) } }}{{\left| {m^{'} } \right|^{\alpha  } }}} \tilde{f}(k,t) dk }+\left( {\frac{1}{{2\pi }}} \right)\sum\limits_{m ^{'}=  1 ,m^{'} \ne 0}^{+\infty }f_{n + m^{'}}^* f_{n } \int\limits_{ - \infty }^{ + \infty } { {\frac{{Je^{ik(m^{'}+n) } }}{{\left| {m^{'} } \right|^{\alpha  } }}} \tilde{f}(k,t) dk}\nonumber \\
&=&b^{\alpha}\int\limits_{ b }^{ + \infty }dy\frac{\left|f(x - y,t)\right|^2  f(x,t)+\left|f(x + y,t)\right|^2f(x,t)}{{\left|y \right|^{\alpha  } }}.
 \end{eqnarray}
 We get  the following  fractional integro-differential cubic nonlinear equation given by
 \begin{eqnarray}\label{eq66}
i\hbar \dot f&=&F a_\alpha  \frac{{\partial ^{\alpha-1} 
f(x,t)}}{{\partial \left| x \right|^{\alpha-1}   }} 
 - Ga_\alpha f^2 \frac{{\partial^{\alpha-1}   f^* }}{{\partial \left| x \right|^{\alpha-1}   }}
 -Ga_\alpha  \left| {f(x,t)} \right|^2 \frac{{\partial ^{\alpha-1} 
f(x,t)}}{{\partial \left| x \right|^{\alpha-1}   }}+5\hat J_{\alpha  }
(0)f\left| f \right|^2 - T f + 2I\left| f \right|^2f \nonumber\\
&&+G\left[ {b^{\alpha}\int\limits_{ b }^{ + \infty }dy\frac{\left|f(x - y,t)\right|^2  f(x - y,t)+
\left|f(x + y,t)\right|^2f(x + y,t)}{{\left|y \right|^{\alpha  } }}}\right]\nonumber\\
&& +G\left[{4b^{\alpha}\int\limits_{ b }^{ + \infty }dy\frac{\left|f(x - y,t)\right|^2  f(x,t)+\left|f(x + y,t)\right|^2f(x,t)}{{\left|y \right|^{\alpha  } }}
 } \right]. \nonumber\\ 
\end{eqnarray}
Next, while introducing the following Ansatz:
\begin{equation}\label{eq67}
f(x,t) = \psi (x,t)\, e^{-i{{2\varepsilon ^2 A t}}/{\hbar }}
\end{equation}
We get  the integro-fractional differential nonlinear Schr\"odinger equation given by Eq.~(\ref{eq23}).

\end{document}